\documentclass[a4paper,11pt]{article}
\usepackage{braket}
\usepackage{pos}
\usepackage{multirow}
\usepackage{amsmath}
\usepackage{caption,subcaption}

\newcommand{\blue}{\color[rgb]{0.00500000,0.06600,1.000} }
\newcommand{\olive}{\color[rgb]{0.30500000,0.56600,0.129000} }
\newcommand{\red}{\color[rgb]{1.00000000,0.06600,0.129000} }

\def\eps{\epsilon}

\title{Regge poles and cuts and the Lipatov vertex}

\author[a,b]{Samuel Abreu}
\author[a,c,e]{Giulio Falcioni}
\author*[a]{Einan Gardi}
\author[d]{Calum Milloy}
\author[d]{Leonardo Vernazza}

\affiliation[a]{Higgs Centre for Theoretical Physics, School of Physics and Astronomy,\\The University of Edinburgh, Edinburgh EH9 3FD, Scotland, UK}
\affiliation[a]{Higgs Centre for Theoretical Physics, School of Physics and Astronomy,\\ The University of Edinburgh, Edinburgh EH9 3FD, Scotland, UK}
\affiliation[b]{Theoretical Physics Department, CERN, Geneva 1211, Switzerland}
\affiliation[c]{Physik-Institut, Universit\"{a}t Z\"{u}rich, Winterthurerstrasse 190, 8057 Z\"{u}rich, Switzerland}
\affiliation[d]{INFN, Sezione di Torino, Via Pietro Giuria 1, I-10125 Torino, Italy}
\affiliation[e]{Dipartimento di Fisica, Universit\`{a} di Torino, Via Pietro Giuria 1, I-10125 Torino, Italy}

\emailAdd{Samuel.Abreu@ed.ac.uk}
\emailAdd{Giulio.Falcioni@ed.ac.uk}
\emailAdd{Einan.Gardi@ed.ac.uk}
\emailAdd{Calum.Milloy@to.infn.it}
\emailAdd{Leonardo.Vernazza@to.infn.it}

\abstract{Scattering amplitudes in the high-energy limit can be described in terms of their singularity structure in the complex angular momentum plane, consisting of Regge poles and cuts. In QCD, gluon Reggeization has long been understood as a manifestation of a Regge pole, but until recently Reggeization violation remained largely obscure. New methods, based on iterative solution of rapidity evolution equations, facilitate direct computation of components of the amplitude which are mediated by multi-Reggeon exchange, a manifestation of Regge cuts. Upon disentangling the Regge cut from the pole we are now able to extract the pole parameters from state-of-the-art fixed-order computations (3 loops) and make predictions regarding certain components of the amplitude to higher loop orders. In this talk I review the key ideas which led to this progress, describe where we stand in exploring the structure of $2\to 2$ and $2\to 3$  amplitudes in the (multi-) Regge limit, and comment on the interplay between this research and the study of infrared factorization.
}

\FullConference{Loops and Legs in Quantum Field Theory (LL2024)\\
 14-19, April, 2024\\
Wittenberg, Germany\\}

\begin{document}
\maketitle

\section{Introduction}
The Regge limit has been a prominent topic in the study of scattering amplitudes since the days of Regge theory, and remains of key interest in unravelling QCD dynamics. While the subject has fascinating non-perturbative aspects, it is tremendously rich already at the perturbative level. 
Recent progress in fixed-order computations of $2\to 2$ amplitudes, which are now known to three loops~\cite{Caola:2022dfa,Caola:2021rqz,Caola:2021izf}, and $2\to 3$ amplitudes, which have recently been computed in full to two loops~\cite{Agarwal:2023suw,DeLaurentis:2024arp,DeLaurentis:2023nss,DeLaurentis:2023izi}, combined with the development of an effective description of high-energy scattering in terms of Reggeon fields using rapidity evolution equations~\cite{Caron-Huot:2013fea,Caron-Huot:2017fxr,Caron-Huot:2017zfo,Gardi:2019pmk,Caron-Huot:2020grv,Falcioni:2020lvv,Falcioni:2021buo,Falcioni:2021dgr}, has led to significant progress: it is now possible to understand in detail and determine explicitly three towers of logarithms involving one, two and three Reggeon exchange. This allows us to make all-order predictions of higher-loop corrections with a given logarithmic accuracy, make direct contact with the notions of Regge pole and Regge cut in $2\to 2$ scattering, and use the Regge-pole parameters in $2\to 3$ scattering to determine the Lipatov vertex at two loops. This information can then be further used to predict higher-point QCD amplitudes in multi-Regge kinematics, where all emissions are ordered in rapidity, and step towards computing third-order corrections to rapidity evolution equations.

Over the past 15 years there has been major progress in understanding the multi-Regge limit at any multiplicity in planar ${\cal N}=4$ super-Yang-Mills theory (see e.g.~\cite{Bartels:2009vkz,Bartels:2008ce,Bern:2005iz,DelDuca:2019tur}). One of the key features of the planar theory is that 4-point and 5-point amplitudes are entirely free of Regge cuts.  Higher multiplicity amplitudes do develop Regge cuts, but only in particular kinematic regions, which can be reached by analytic continuation. In contrast, the full, non-planar theory --- be it super-Yang-Mills or QCD --- has a much richer structure in the Regge limit, already at low multiplicities. It is this structure which we aim to explore. 

In this talk I will first review the notion of Regge factorization in the context of $2\to 2$ QCD scattering (Section~\ref{Sec:Reggeization}), and explain how it is violated starting from next-to-next-to-leading logarithmic (NNLL) accuracy due to multiple-Reggeon exchange. Next, in Section~\ref{Sec:B-JIMWLK_and_MR}, I will briefly outline how the use of non-linear rapidity evolution equations~\cite{Caron-Huot:2017fxr,Caron-Huot:2017zfo,Gardi:2019pmk,Caron-Huot:2020grv,Falcioni:2020lvv,Falcioni:2021buo,Falcioni:2021dgr}, following the work of Simon Caron-Huot~\cite{Caron-Huot:2013fea}, has transformed our ability to make predictions regarding multi-Reggeon exchange contributions to partonic amplitudes. In Section~\ref{Sec:Pole-cut} I will show an application of these techniques and explain how Regge-pole and Regge-cut contributions can be disentangled at NNLL. In Section~\ref{Sec:Lipatov_Vertex} I will turn to discuss $2\to 3$ scattering and present new results for multi-Reggeon exchange we obtained using the techniques of Section~\ref{Sec:B-JIMWLK_and_MR}, which facilitate extracting the Lipatov vertex at two loops from recent amplitude computations.

\section{$2\to 2$ amplitudes: Reggeization and its breaking }
\label{Sec:Reggeization}

\begin{figure}[t]
\centering
\includegraphics[width=0.58\textwidth]{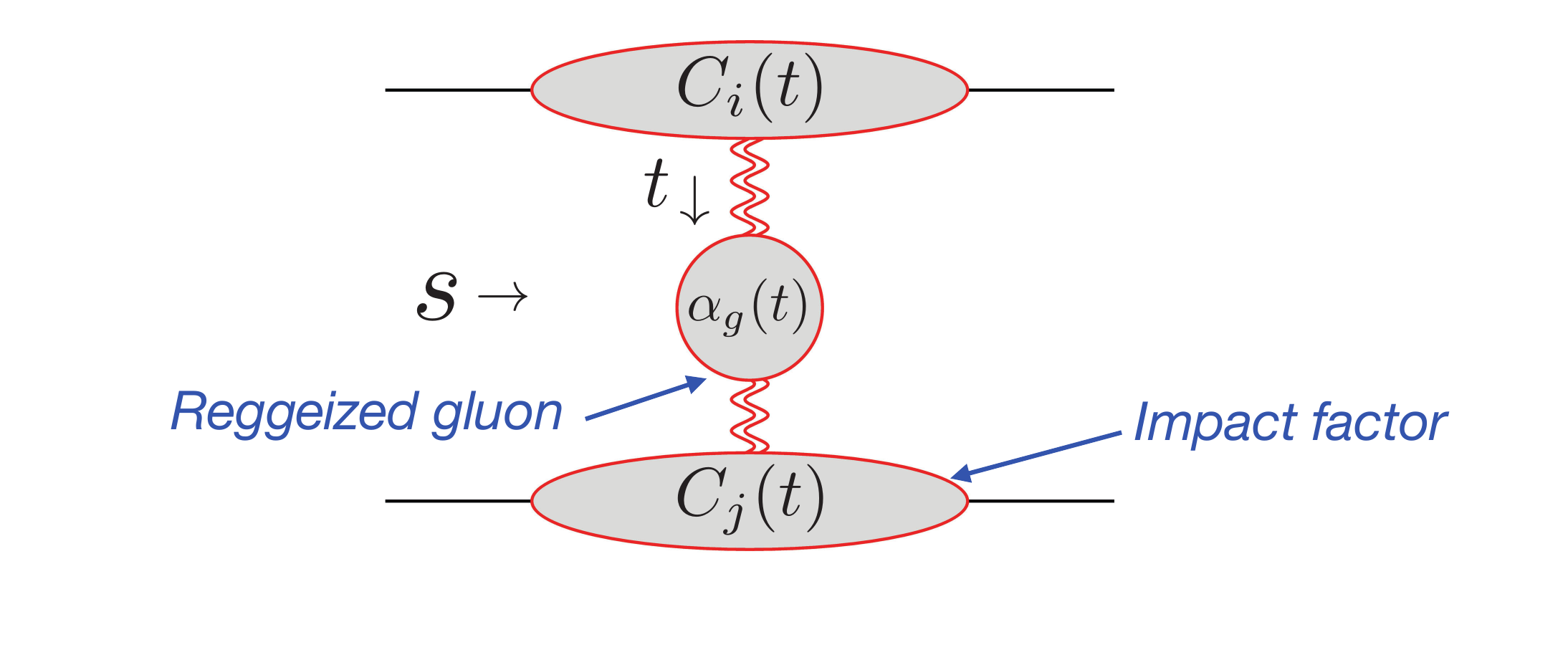}
\vspace*{-20pt} 
\caption{Regge (pole) factorization in $2\to 2$ scattering. Here the wavy line represents a Reggeized gluon exchanged in the $t$ channel, while the scattered particles ($i$ and $j$) may be quarks or gluons. The impact factors $C_{i/j}$ represent the effective vertex for single Reggeon emission.}
\label{fig:factorization}
\end{figure}

QCD scattering amplitudes simplify drastically in the Regge limit. The simplest example is $2\to 2$ scattering, $gg\to gg$, $qq\to qq$ or $qg\to qg$, in the limit $s\gg -t\gg\Lambda_{\rm QCD}$, where the leading power in the $-t/s\to 0$ expansion is given by the helicity-conserving configuration, and is governed (at tree level) by a 
$t$-channel gluon exchange, as illustrated in figure~\ref{fig:factorization}. The dominance of gluon exchange is a manifestation of the general prediction from Regge theory~\cite{Collins:1977jy,White:2019ggo}, that scattering amplitudes behave in this limit as $s^\ell$, where $\ell$ is the spin of the particle exchanged in the $t$ channel. 

More profoundly, power-like growth of partonic QCD  amplitudes persists to loop level: at leading-logarithmic (LL) accuracy the amplitude behaves as $(s/(-t))^{1+\alpha_g(t)}$ where $\alpha_g(t)$ is the gluon Regge trajectory. This amounts to all-order exponentiation of rapidity logarithms, $\ln(s/(-t))$~\cite{Lipatov:1976zz}.
This, so-called \emph{gluon Reggeization} phenomenon, is a manifestation of a Regge pole in the complex angular momentum plane, corresponding to a Reggeized gluon exchange in the $t$ channel.   

At one loop the gluon Regge trajectory in dimensional regularization ($D=4-2\epsilon$) is:
\begin{align}
\label{one_loop_traj}
\alpha_g(t)=
-\alpha_s {\mathbf T}_t^2 (\mu^2)^\epsilon \int\frac{d^{2-2\epsilon}k_\perp}{(2\pi)^{2-2\epsilon}} \frac{q_\perp^2}{k_\perp^2 (q-k)_{\perp}^2}
=
\frac{\alpha_s}{\pi} {\mathbf T}_t^2  \left(\frac{\mu^2}{-t}\right)^{\epsilon}\frac{r_{\Gamma}}{2\epsilon}
+{\cal O}(\alpha_s^2)\,,
\end{align}
where~${\bf T}_t^2$ is the quadratic Casimir operator representing the colour charge carried across the $t$ channel, and where
\begin{align}
\label{rGamma}
r_{\Gamma} \equiv e^{\epsilon \gamma_E} \frac{\Gamma^2(1-\epsilon)\Gamma(1+\epsilon)}{\Gamma(1-2\epsilon)} = 1 - \frac{\zeta_2}{2} \epsilon^2 - \frac{7\zeta_3}{3} \epsilon^3 + \ldots\,.
\end{align}
Note that having extracted the large logarithm, $\ln(s/(-t))$, the remaining loop integration in (\ref{one_loop_traj}) is in the transverse momentum plane. We will see below that this is a hallmark of the effective description of the high-energy limit using Reggeons. Recently, the gluon Regge trajectory $\alpha_g(t)$ has been determined to three-loop order in QCD~\cite{Falcioni:2021buo,Falcioni:2021dgr,Caola:2021izf,Caola:2021rqz}. This will be briefly discussed in Section~\ref{Sec:Pole-cut}.

In Regge theory the asymptotic behaviour of the amplitude at large $s$ is directly linked to 
their analytic structure
in the complex angular momentum plane. This formalism is based on expressing the $t$-channel amplitude as a sum over states with a
given angular momentum~$\ell$, and analytically continuing to the $s$ channel. This procedure requires separating between even and odd values of~$\ell$, leading to amplitudes of even and odd signature, respectively, which are symmetric (antisymmetric) with respect to the interchange of the $s$ and $u$ Mandelstam invariants:
\begin{equation}
    {\cal M}^{(\pm)}(s,t) = \tfrac12\Big( {\cal M}(s,t) \pm {\cal M}(-s-t,t) \Big)\,.
\end{equation}
Interestingly, signature can be associated with reality properties (see~\cite{Caron-Huot:2017fxr} and refs. therein): using a Mellin transform of the spectral representation of the amplitude one finds that the even and odd amplitudes, respectively, can be written as 
\begin{equation}
\label{Mplusminus}
   {\cal M}^{(+)}(s,t) = i \,\int_{\gamma- i\infty}^{\gamma+i\infty} 
 \frac{dj}{2\sin (\pi j/2)} \, a^{(+)}_j(t) \, e^{jL}\,, \quad 
{\cal M}^{(-)}(s,t) =  \int_{\gamma- i\infty}^{\gamma+i\infty} 
 \frac{dj}{2\cos (\pi j/2)}\, a^{(-)}_j(t) \, e^{jL}\,,
 \end{equation}
where $L \equiv  \log\left|\frac{s}{t}\right| -i\frac{\pi}{2}= \frac12\left(\log\frac{-s-i0}{-t}+\log\frac{-u-i0}{-t}\right)$. 

The reality property of the spectral density functions implies $\left(a_{j^*}^{\pm}(t) \right)^*=a_j^{\pm}(t)$, hence eq.~(\ref{Mplusminus}) shows that the expansion coefficients of ${\cal M}^{(+)}(s,t)$ in powers of $L$ are purely imaginary, while those of ${\cal M}^{(-)}(s,t)$ are purely real. The asymptotic behaviour of ${\cal M}^{(+)}$ and ${\cal M}^{(-)}$ are therefore distinct and have been studied separately. 

Fundamental to our understanding of the behaviour of QCD amplitudes in the Regge limit is the notion of colour flow. Since a Reggeon, like a gluon, transforms according to the antisymmetric octet representation of the colour group, it is convenient to decompose colour-dressed amplitudes in $t$-channel colour basis (see e.g.~\cite{DelDuca:2011ae,Caron-Huot:2017fxr}). While tree-level scattering is a pure octet exchange, loop corrections populate other components in the $t$-channel basis. Given the classification into odd and even signature (symmetry under $s\leftrightarrow u$), it is natural to classify the colour representations according to their symmetry under a similar swap between initial and final state particles, on either the target side or the projectile side.  
The correspondence with signature for gluon-gluon scattering is simple: Bose symmetry implies that ${\cal M}^{(-)}$ consists of odd colour representations ($8_a \oplus (10\oplus\overline{10}))$, while ${\cal M}^{(+)}$ even ones ($0\oplus 1\oplus 8_s\oplus 27$).

While the aforementioned Reggeization phenomenon seen at LL accuracy corresponds to ${\cal M}^{(-)}$ having a simple Regge pole, more generally, both ${\cal M}^{(+)}$ and ${\cal M}^{(-)}$  also feature Regge cuts: 
\begin{equation}
a_j^{(\pm)}(t) \simeq \frac{1}{(j-1-\alpha(t))^{1+\beta(t)}}\,\, \Rightarrow \,\, 
{\cal M}^{(\pm)}(s,t)|_{\rm Regge\, cut} \simeq 
\frac{\pi }{\sin \frac{\pi\, \alpha(t)}{2}}  \frac{s}{t} 
\, \frac{1}{\Gamma\left(1+\beta(t)\right)} L^{\beta(t)}  
\, e^{L\, \alpha(t)}\,, 
\end{equation}
where subleading powers and subleading logarithms have been neglected. 

In QCD, since the Reggeon has odd signature under the exchange of the initial with the final states, ${\cal M}^{(-)}$ receives contributions from the exchange of an odd number of Reggeons, while ${\cal M}^{(+)}$ an even number.
The signature-even amplitude ${\cal M}^{(+)}(s,t)$ begins at next-to-leading logarithmic accuracy (NLL), and is generated by the $t$-channel exchange of \emph{two} Reggeized gluons, which form a Regge cut. This tower of logarithms has been studied in detail in refs.~\cite{Caron-Huot:2017zfo,Caron-Huot:2020grv,Gardi:2019pmk}, where explicit computations have been carried out up to very high loop orders. 
This was possible thanks to the fact that the two-Reggeon wavefunction admits the BFKL equation~\cite{Fadin:1975cb,Kuraev:1977fs,Kuraev:1976ge,Balitsky:1978ic}. 

In this talk we focus on the signature-odd sector, where a single Reggeon exchange continues to dictate the structure of higher-loop corrections at NLL~\cite{Fadin:1995xg,Fadin:1995km,Fadin:2015zea}. ${\cal M}^{(-)}(s,t)$ at NLL is thus still described by a simple Regge pole. The octet component of the  amplitude then factorises as in figure~\ref{fig:factorization}, 
\begin{equation}
\label{eq-pole-fact}
{\cal M}_{ij\to ij}^{(-)\,{\rm pole}} 
=
C_i(t)\,e^{\alpha_g(t)\,C_A\,L} \,C_j(t) {\cal M}^{\text{tree}}_{ij\to ij}  \,,
\end{equation}
so NLL corrections manifest themselves both as two-loop corrections to the gluon Regge trajectory $\alpha_g(t)$, or as one-loop corrections to the impact factors $C_{i/j}(t)$ (see figure~\ref{fig:factorization}). 
In contrast, at next-to-next-to-leading logarithmic accuracy (NNLL), in addition to single Reggeon exchange, new contributions arise from \emph{triple} Reggeon exchange. 
These effects, which we shall briefly review in Section~\ref{Sec:Pole-cut}, give rise to a Regge cut, and break the Regge-pole factorization structure of eq.~(\ref{eq-pole-fact}).  

The breakdown of Regge-pole factorization at NNLL accuracy can be detected already at two loops, the first order at which three gluons can be exchanged in the $t$ channel. 
Ref.~\cite{DelDuca:2001gu}
has shown that the three two-loop amplitudes,
$gg\to gg$, $qq\to qq$ and $qg\to qg$ 
(in the octet channel) cannot be simultaneously consistent with the  factorization formula (\ref{eq-pole-fact}). Non-factorizable terms in ${\cal M}_{ij\to ij}^{(-)}$ have then been singled out in ref.~\cite{DelDuca:2013dsa,DelDuca:2013ara,DelDuca:2014cya} based on the infrared singularity structure. Ultimately, identifying contributions that originate in multiple Reggeon exchange requires an independent computation of these effects in terms of Reggeons.  A framework to do that has been developed in recent years~\cite{Caron-Huot:2017fxr,Falcioni:2020lvv,Falcioni:2021buo,Falcioni:2021dgr}, based on the pioneering paper by Simon Caron-Huot~\cite{Caron-Huot:2013fea}, which makes use of non-linear rapidity evolution equations to describe partonic scattering in the high-energy limit. We briefly introduce the key idea in the next section. We comment that an alternative Feynman-diagram-based approach has been developed by Fadin and Lipatov~\cite{Fadin:2017nka,Fadin:2024hbe,Fadin:2023aen}.

\section{Multi-Reggeon exchange computations from rapidity evolution} 
\label{Sec:B-JIMWLK_and_MR}

In the shockwave formalism~\cite{Balitsky:1995ub,Jalilian-Marian:1996mkd,JalilianMarian:1996xn,JalilianMarian:1997gr,Caron-Huot:2013fea,Caron-Huot:2017fxr,Falcioni:2021buo} 
one describes the projectile as a product 
of (an indefinite number of) Wilson-line 
operators, $U({z_1}_{\perp})U({z_2}_{\perp})\ldots$, 
each of which extends over the infinite $+$ 
lightcone direction and located as a distinct 
transverse position $z_{\perp}$:
\begin{equation}
\label{Udef}
U(z_{\perp})={\cal P}\,\exp\left[ \int_{-\infty}
^{\infty}T^a A_{+}^a(x^{+} ,x^{-}=0, z_\perp) 
dx^+ \right]\,.
\end{equation}
The Wilson lines (\ref{Udef}) have 
rapidity divergences, so the operators 
$U({z_1}_{\perp})U({z_2}_{\perp})\ldots$ 
admit evolution equations in the rapidity $\eta$, taking the form 
\begin{equation}
\label{H_acting_on_UUU}
-\frac{d}{d\eta} \left[U({z_1}_{\perp})U({z_2}_{\perp})\ldots\right] 
= H \left[U({z_1}_{\perp})U({z_2}_{\perp})\ldots\right]\,,
\end{equation}
known as the Balitsky-JIMWLK 
equations~\cite{Balitsky:1995ub,Jalilian-Marian:1996mkd,JalilianMarian:1996xn,JalilianMarian:1997gr}. 
The leading-order Hamiltonian is 
\begin{equation}
\label{B-JIMWLK-H}
H=\frac{\alpha_s}{2\pi^2} \mu^{2\epsilon}\int 
d^{2{-}2\eps}z_0
\frac{{ z}_{0i} \cdot { z}_{0j}}{
\big({ z}_{0i}^2 { z}_{0j}^2\big)^{1-\epsilon}} 
\left(T_{i,L}^a T_{j,L}^a +T_{i,R}^a T_{j,R}^a 
- U_{\rm adj}^{ab}({ z}_0) 
(T_{i,L}^a T_{j,R}^b +T_{j,L}^a T_{i,R}^b )\right)\,,
\end{equation}
where $z_{0i}\equiv z_0-z_i$. Here we used functional derivative 
operators, which generate 
colour rotation:
\begin{equation}
\label{Tder}
T_{i,L}^a \equiv T^aU({ z}_i)\frac{\delta}{\delta U({ z}_i)},
\qquad 
T_{i,R}^a \equiv U({ z}_i)T^a\frac{\delta}{\delta U({ z}_i)}\,.
\end{equation}
Crucially, the Hamiltonian acts exclusively in the $2-2\eps$ dimensional
transverse plane. The action of the Hamiltonian $H$ on the product 
of Wilson lines involves an additional Wilson line 
in the adjoint representation, $U_{\rm adj}^{ab}({ z}_0)$,
generated through the interaction with the target 
shockwave. Thus, this system of equations is 
non-linear and each iteration of the evolution 
involves an increasing number of Wilson lines. 

The non-linear nature of the equation has been shown to be essential in capturing the transition into the high-gluon density saturation regime in hadronic or heavy-ion scattering. The context in which we use the equation is however, different: we are interested in the perturbative regime where fields are weak, and each Wilson $U(z_{\perp})$ 
is close to unity. The dynamics is then best 
described in terms of $W$, defined 
via~\cite{Caron-Huot:2013fea}
\begin{equation}
U(z_{\perp}) = e^{ig_s {\bf T}^a W^a(z_{\perp})}\,.
\end{equation}
A key observation~\cite{Caron-Huot:2013fea} is that  field~$W$ can be identified as sourcing \emph{a 
single Reggeon}, which is exchanged in the $t$ channel. 
Wilson lines~$U$ (and products thereof) can 
be thus described perturbatively, order by 
order in $g_s$ as an expansion in $W$ fields, 
all at the same transverse position,  
\begin{eqnarray}
     \label{UexpansionW} 
U= e^{ig_s \,W^a {\bf T}^a} = {\bf 1}+ ig_s \, W^a \, {\bf T}^a 
- \frac{g_s^2}{2} W^a W^b \, {\bf T}^a {\bf T}^b
-i\frac{g^3_s}{6}W^a W^b W^c\, {\bf T}^a {\bf T}^b {\bf T}^c+\cdots  
\end{eqnarray}
where $n$ fields $W$ source $n$ 
Reggeons.

Next one needs to establish how the Hamiltonian (\ref{H_acting_on_UUU}) acts on products of Reggeon $W$ fields. To this end one must first expand the colour rotation operators in (\ref{Tder}) in $W$ fields. This yields a matrix Hamiltonian of the form~\cite{Caron-Huot:2017fxr,Falcioni:2021buo} 
\begin{align}
\label{H_mat_form}
\begin{split}
\def\bra#1{\langle#1|}
\def\ket#1{|#1\rangle}
H  \left( 
\begin{array}{c}
  W      \\  WW     \\  \!\!\!WWW\!\!\!  \\  \cdots
\end{array}
\right) \equiv
\left(
\begin{array}{cccc}
 H_{1{\to}1} & 0  & H_{3{\to}1} & \ldots \\
 0 & H_{2{\to}2}  & 0  &  \ldots \\
 H_{1{\to}3} & 0  & H_{3{\to}3} & \ldots\\
 \cdots & \cdots  & \cdots & \cdots\\
\end{array}
\right)\left(
\begin{array}{c}
  W      \\  WW     \\ \!\!\! WWW\!\!\!  \\  \cdots\end{array}
\right)
\sim
\left(
\begin{array}{cccc}
 g_s^2 & 0  & g_s^4 & \ldots \\
 0 & g_s^2  & 0  &  \ldots \\
 g_s^4 & 0  & g_s^2 & \ldots\\
 \cdots & \cdots  & \cdots & \cdots\\
\end{array}
\right) \left(
\begin{array}{c}
  W      \\  WW     \\  \!\!\!WWW\!\!\!  \\  \cdots\end{array}
\right)    
\end{split}
\end{align}
where the first matrix expression defines the components 
of the Hamiltonian $H_{k\to m}$ mediating 
between a state with $k$ Reggeons and one 
with $m$ Reggeons, where transitions 
between odd and even number of Reggeons 
are forbidden. The second matrix expression
indicates the order in $g_s$ at which 
each such component of the Hamiltonian 
begins. In this way the perturbative 
expansion of the Wilson lines in effect 
linearises the non-linear evolution, and 
upon working to a given order in the 
coupling, the system involves a restricted 
number of Reggeons.

To construct a scattering amplitude we need to express the scattering partons in terms of states with a definite number of Reggeons. 
At leading order
\begin{align}
\label{eq:psi3}
\begin{split}
\ket{\psi_{i}}&=\sum_{n=1}^{\infty} \ket{\psi_{i,n}}=
ig_s\,{\bf T}_i^a \, W^a(p)|0\rangle
- \frac{g_s^2}{2} {\bf T}_i^a{\bf T}_i^b 
\int 
\frac{d^{2-2\eps} q}{(2\pi)^{2-2\eps}} \, 
W^a(q)W^b(p{-}q)|0\rangle
\\
&
-\, \frac{ig_s^3}{6} {\bf T}_i^a{\bf T}_i^b{\bf T}_i^c 
\int 
\frac{d^{2-2\eps} q_1}{(2\pi)^{2-2\eps}} \, 
\frac{d^{2-2\eps} q_2}{(2\pi)^{2-2\eps}} \,
W^a(q_1)W^b(q_2)W^c(p{-}q_1{-}q_2)|0\rangle +\ldots 
\end{split}
\end{align}
where $n$ counts the number of Reggeon fields $W$.
The $2\to 2$ amplitude in the high-energy limit is then obtained by evolving the target to the rapidity of the projectile or vice versa, 
fixing the rapidity difference $\eta$ to be $L$, defined below eq.~(\ref{Mplusminus}).
The amplitudes then takes the form
\begin{equation}
\label{2to2amplitudeUsingH}
    \frac{i}{2s}
{\cal M}_{ij\to ij} =
Z_i Z_j C_i^{(0)} C_j^{(0)} \braket{\psi_j|e^{-H L}|\psi_i}\,, 
\end{equation}
where $C_{i/j}^{(0)}$ are the leading-order impact factor and $Z_{i/j}$ restore the collinear singularities. 

The predictive power of this approach is due to the fact that towers of energy logarithms $L$ can be obtained by expanding the evolution operator $e^{-H L}$ and then repeatedly applying the Hamiltonian on the scattered parton states. At a given logarithmic accuracy, there is an upper limit on the number of Reggeons $W$ that can be exchanged, which in turn restricts the relevant components in the Hamiltonian (\ref{H_mat_form}).  
Importantly, the action of the Hamiltonian is strictly in the transverse space (Euclidean, $2-2\epsilon$ dimensional space), so the integrals do not require any additional regulator. 
This may be contrasted with the alternative effective descriptions of the high-energy limit using Glauber Soft-Collinear Effective Theory~\cite{Rothstein:2016bsq,Moult:2022lfy,Gao:2024qsg}, where integrals do still involve the longitudinal degrees of freedom, and require rapidity regulators. In the next two sections we will present applications of the effective Reggeon formalism, first to $2\to 2$ scattering and then to $2\to 3$ scattering.

\section{Disentangling Regge pole and cut at NNLL}
\label{Sec:Pole-cut}

Let us now focus on the NNLL tower of corrections to the signature-odd $2\to 2$ amplitude following refs.~\cite{Caron-Huot:2017fxr,Falcioni:2020lvv,Falcioni:2021buo, Falcioni:2021dgr}. As discussed in section~\ref{Sec:Reggeization}, this is the first logarithmic order where a Regge cut appears, in addition to the Regge pole. As anticipated there, the cut reflects the exchange of three Reggeized gluons, which first contributes at two loops. The formalism of section~\ref{Sec:B-JIMWLK_and_MR} allows us to characterize and compute explicitly all multiple-Reggeon exchange contributions. At NNLL accuracy these are limited to three Reggeon exchanges and their evolution, as well as as their mixing with a single Reggeon. 

The NNLL tower of corrections was first studied using this formalism in ref.~\cite{Caron-Huot:2017fxr} 
 through three loops. Subsequently, its all-order structure has been elucidated in refs.~\cite{Falcioni:2020lvv,Falcioni:2021buo, Falcioni:2021dgr}, which also extended the explicit computation to four loops. Following this work, in order to compactly present the structure of the NNLL tower it is useful to define a reduced amplitude by $\hat{\cal M}_{ij\to ij} \equiv 
\left(Z_i  Z_j \right)^{-1} 
\,e^{-\,{\bf T}_t^2\, \alpha_g(t) \, L} \, {\cal M}_{ij\to ij}$, effectively removing the effect of a single Reggeon  rapidity evolution. Using (\ref{2to2amplitudeUsingH}) one has $\frac{i}{2s}\hat{\cal M} = e^{-\,{\bf T}_t^2\, \alpha_g(t) \, L}
\braket{\psi_j|e^{-H L}|\psi_i} \equiv 
\braket{\psi_j|e^{-\hat H L}|\psi_i}$, where the reduced Hamiltonian is defined by 
${\hat{H}}_{k\to k+2n} = H_{k\to k+2n} +\delta_{n0}{\bf T}_t^2\alpha_g(t)$. Expressing both the target and the projectile as an expansion in terms of states of a definite number of Reggeon field $W$ as in (\ref{eq:psi3}), and considering the evolution in rapidity through repeated application of the 
Hamiltonian~(\ref{H_mat_form}), one finds~\cite{Falcioni:2020lvv} that the entire tower of NNLL corrections is given by 
\begin{align}
\label{NNLLtower}
\begin{split}
    \frac{i}{2s}\hat{\cal M}^{(-),\text{NNLL}}_{ij\to ij}\!=\left(\frac{\alpha_s}{\pi}\right)^2\Bigg\{&
\braket{j_{1}|i_{1}}^{\text{NNLO}}+
r_{\Gamma}^2\pi^2 
\sum_{k=0}^{\infty}\frac{(-X)^k}{k!} \Bigg[\braket{j_{3}|\hat{H}_{3\to3}^k|i_{3}}  \\
&+\Theta(k\geq 1)
\left[
\braket{j_{1}|\hat{H}_{3\to1}\hat{H}_{3\to3}^{k-1}|i_{3}}+
\braket{j_{3}|\hat{H}_{3\to3}^{k-1}\hat{H}_{1\to3}|i_{1}}\right]\\
&      +\Theta(k\geq2)\braket{j_{1}|\hat{H}_{3\to1}\hat{H}_{3\to3}^{k-2}\hat{H}_{1\to3}|i_{1}}\Bigg]^{\text{LO}}\Bigg\},
\end{split}
\end{align}
where we defined the $n$ Reggeon states of (\ref{eq:psi3}) as $\ket{\psi_{i,n}}\equiv (r_\Gamma\alpha_s)^{(n-1)/2} \,\ket{i_n}$ and the expansion parameter $X\equiv r_\Gamma\alpha_s L/\pi$. Here $L$ is the symmetric logarithm of eq.~(\ref{Mplusminus}) and $r_\Gamma$ was defined in eq.~(\ref{rGamma}).

\begin{figure}[t]
\centering
\begin{subfigure}[t]{0.33\textwidth}
\centering
\includegraphics[height=3.25cm]{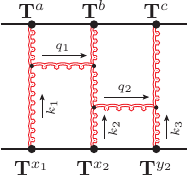}
\hspace*{-20pt}\caption{$3\to 3$ Reggeon exchange}
\label{fig:MR:4loops_3to3}
\quad
\end{subfigure}
\begin{subfigure}[t]{0.3\textwidth}
\centering
\includegraphics[height=3.2cm]{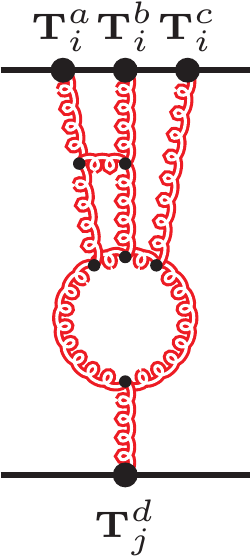}
\caption{$1\to 3$ Reggeon exchange}
\label{fig:MR:4loops_1to3}
\quad
\end{subfigure}
\begin{subfigure}[t]{0.33\textwidth}
\centering
\includegraphics[height=3cm]{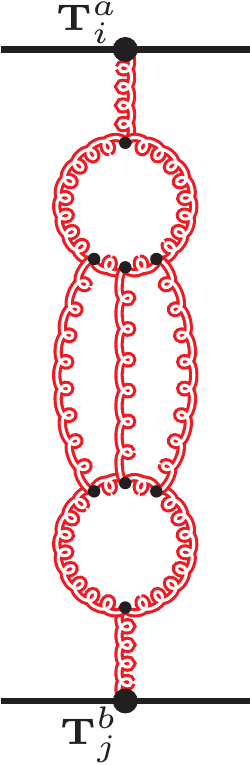}\quad
\includegraphics[height=3cm]{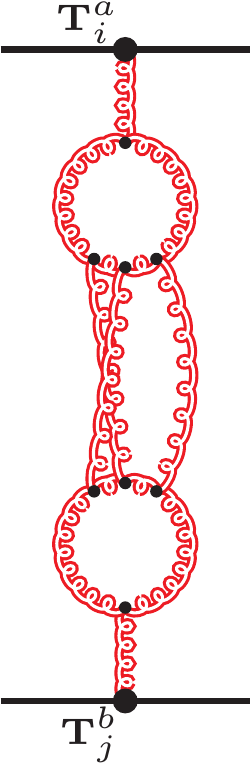}
\caption{$1\to3\to 1$ Reggeon exchanges}
\label{fig:MR:4loops_1to3to1}
\end{subfigure}
\caption{Multi-Reggeon exchange in $2\to 2$ scattering:  four loop examples}
\label{fig:MR:4loops}
\end{figure}

The first term in eq.~(\ref{NNLLtower}), $\braket{j_{1}|i_{1}}^{\text{NNLO}}$, represents the single Reggeon (SR) exchange, which factorizes as in eq.~(\ref{eq-pole-fact}) and figure~\ref{fig:factorization}. All remaining terms in eq.~(\ref{NNLLtower}) involve multiple-Reggeon (MR) exchange. The first of these is the pure three-Reggeon exchange contribution $\braket{j_{3}|\hat{H}_{3\to3}^k|i_{3}}$. For $k=0$ these are the two-loop diagrams associated with triple exchange in the $t$ channel (with any permutation of the colour generators). At higher orders, $k=1,2, \ldots$ these three Reggeons interact through the action of the $3\to 3$ Hamiltonian $\hat{H}_{3\to3}$; each such action adds a rung, building up a three-Reggeon ladder. A four-loop example is shown in figure~\ref{fig:MR:4loops_3to3}. Similarly, the terms in the second line of eq.~(\ref{NNLLtower}) correspond to diagrams where a single Reggeon on the target side evolves into three Reggeons on the projectile side, or vice versa. Such terms contribute starting from three loops. A four loop example of such a term is depicted in figure~\ref{fig:MR:4loops_1to3}. The third line in eq.~(\ref{NNLLtower}) shows the final type of contribution, which shows up starting from four loops; here a single Reggeon evolves into a three-Reggeon state, and then back again into a single Reggeon. Examples of such four-loop diagrams are shown in figure~\ref{fig:MR:4loops_1to3to1}.

A key property of the multi-Reggeon contribution to the signature-odd NNLL tower of corrections (\ref{NNLLtower}) is that it is proportional to $(i\pi)^2$. Another key property is that the contributions to (\ref{NNLLtower}) that are leading in the large-$N_c$ limit are in fact factorizable on a Regge pole\footnote{Note that compatibility with Regge-pole factorization amounts to a 
highly-constrained structure of the corrections in eq.~(\ref{NNLLtower}) in the planar limit: at leading order in $N_c$ the corrections to $gg\to gg$, $qq\to qq$ and $qg\to qg$ must be \emph{identical}, and they must \emph{vanish} at four loops and beyond. Both these properties have been seen to hold through four loops~\cite{Falcioni:2021buo, Falcioni:2021dgr}.}, as in (\ref{eq-pole-fact}). Indeed, it is known that in $2\to 2$ scattering, Regge cut contributions are entirely due to non-planar diagrams and must be subleading at large $N_c$.  
This suggests the following prescription for disentangling between Regge pole and cut contributions at NNLL~\cite{Falcioni:2021buo, Falcioni:2021dgr}:
\begin{align}
\label{pole_cut_separation}
\begin{split}
    {\cal M}^{(-)}_{ij\to ij} &= \underbrace{{\cal M}^{(-)\,{\rm SR}}_{ij\to ij} +
\left.{\cal M}^{(-)\,{\rm MR}}_{ij\to ij}\right\vert_{\text{planar}}} 
+
\left.{\cal M}^{(-)\,{\rm MR}}_{ij\to ij}\right\vert_{\text{nonplanar}} \\
&= \hspace*{40pt} 
{\color[rgb]{0.107,0.3300,0.56250} {\cal M}^{(-)\,{\rm pole}}_{ij\to ij}}
 \hspace*{35pt} + {\color[rgb]{1.0,0.00,0.02} {\cal M}^{(-)\,{\rm cut}}_{ij\to ij}}\,.
 \end{split}
\end{align}

The availability of three-loop results for partonic scattering in QCD~\cite{Caola:2021rqz,Caola:2022dfa,Caola:2021izf} along with the multi-Reggon calculation reported above, allow us
to implement the separation of eq.~(\ref{pole_cut_separation}) and explicitly determine the Regge-pole parameters governing the singature-odd tower of corrections to all orders through NNLL accuracy. The newly determined parameters are the two-loop impact factors through ${\cal O}(\epsilon^2)$ for both quarks and gluons and the three-loop gluon Regge trajectory~\cite{Caola:2021izf,Falcioni:2021dgr}. One may then express the signature-odd amplitude as follows~\cite{Falcioni:2021buo, Falcioni:2021dgr}
\begin{align}
\label{pole_and_cut}
    & \mathcal{M}^{(-)}_{ij\to ij} = 
{\color[rgb]{0.107,0.3300,0.56250}  
C_i(t) 
C_j(t)
\left[\left(\frac{-s}{-t}\right)^{C_A\alpha_g(t)} 
+ \left(\frac{-u}{-t}\right)^{C_A\alpha_g(t)}\right]
\mathcal{M}^{\text{tree}}_{ij\to ij}   
} +{\color[rgb]{1.00,0.00,0.020} \sum_{n=2}^\infty 	a^n L^{n-2} {\cal M}^{(\pm,n,n-2)\,{\rm cut}}}\,,
\end{align}
where the first, factorized term (in blue) represents the Regge pole through NNLL accuracy, while the remaining, factorization-violating term (in red) represents the Regge cut, which originates in the non-planar component of the multi-Regge exchange diagrams of eq.~(\ref{NNLLtower}).
Here the impact factors $C_{i/j}(t) =Z_{i/j}(t)\,\bar D_{i/j}(t)$ where $Z_{i/j}(t)$ capture the collinear singularities and $\bar{D}_{i/j}(t)$ are finite. The remaining singularities in the Regge-pole part are contained in the gluon Regge trajectory, and remarkably, are directly proportional to the integral over the lightlike cusp anomalous dimension, 
${{\color[rgb]{0.107,0.3300,0.56250}  \alpha_g(t) = -\frac14 \int_0^{\mu^2} \frac{d\lambda^2}{\lambda^2} \gamma_K(\alpha_s(\lambda^2)) \,+\,{\cal O}(\epsilon^0)}}$,
as originally observed at two loops in Ref.~\cite{Korchemskaya:1994qp,Korchemskaya:1996je}. Ref.~\cite{Falcioni:2021buo,Falcioni:2021dgr} found that this relation remains true at three loops (and conjecturally beyond), despite the fact that additional singularities are present in the cut contribution in eq.~(\ref{pole_and_cut}). The latter, in turn, are sensitive to quadruple corrections to the dipole formula, namely soft singularities that correlate between four Wilson lines at three loop and beyond~\cite{Falcioni:2021buo}. In this way the calculations done in the Regge limit at four loops, provide a strong constraint on the four-loop soft anomalous dimension~\cite{Falcioni:2021buo,Falcioni:2021dgr}.

\section{Towards determination of the Lipatov vertex from $2\to3$ amplitudes}
\label{Sec:Lipatov_Vertex}

\begin{figure}[htb]
\centering
\begin{subfigure}[t]{0.45\textwidth}
\centering
\includegraphics[width=0.9\textwidth]{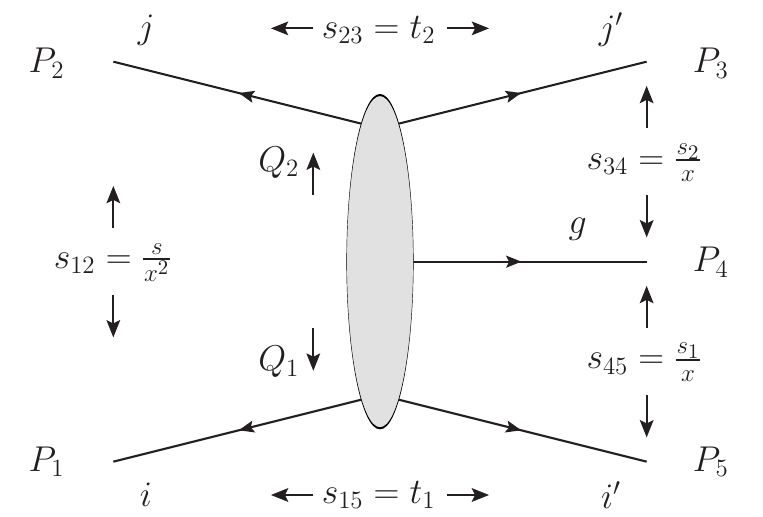}
\caption{Multi-Regge Kinematics (MRK): $x\to 0$}
\label{fig:MRK_5points:kin}
\end{subfigure}
\qquad
\begin{subfigure}[t]{0.3\textwidth}
\centering
\includegraphics[width=0.77\textwidth]{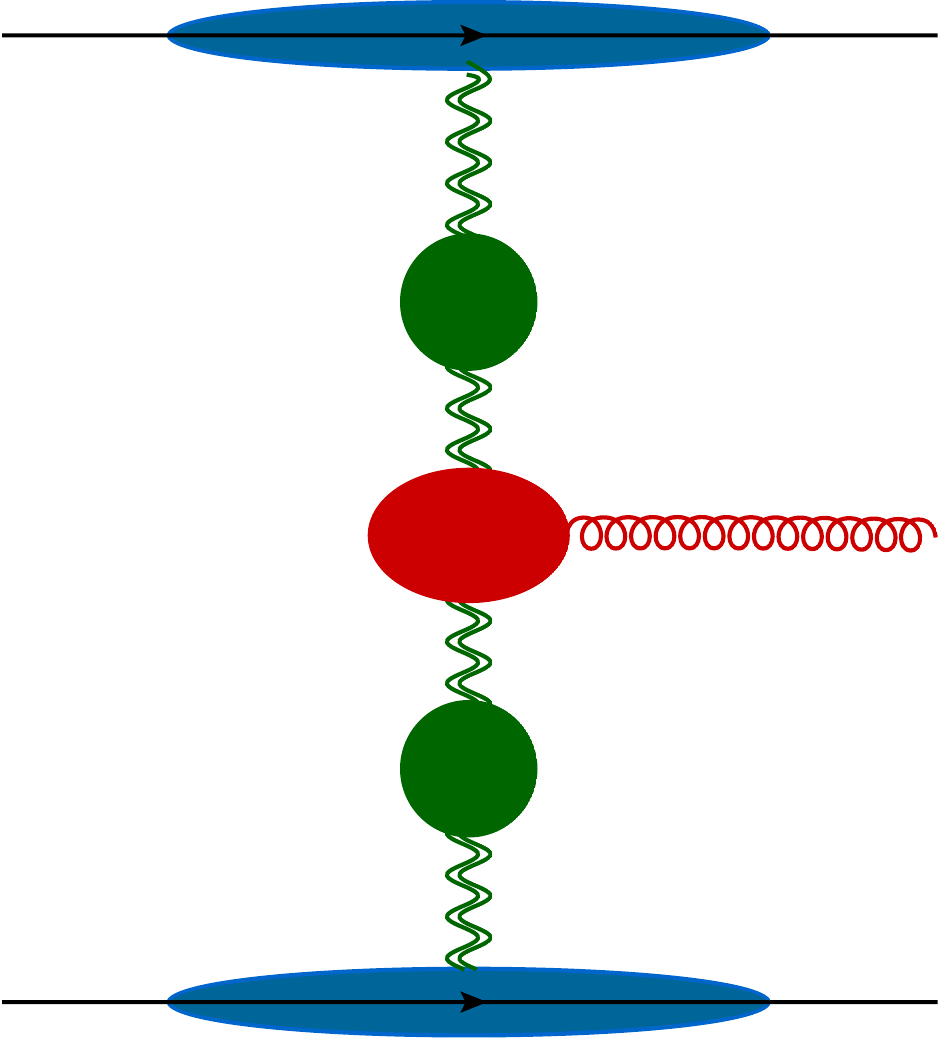}
\caption{Regge-pole factorization}
\label{fig:MRK_5points:fact}
\end{subfigure}
\caption{The MRK limit in $2\to 3$ scattering and the corresponding factorization in the octet-octet channel. We work in the frame where $P_1 = \frac{\sqrt{s_{12}}}{2}{(1,0,0,-1)}$,  
and $P_2 = \frac{\sqrt{s_{12}}}{2}{(1,0,0,1)}$. }
\label{fig:MRK_5points}
\end{figure}

Gluon Reggeization applies, and holds to NLL accuracy~\cite{Fadin:2006bj}, not only for $2\to2$ scattering, but also to higher-point amplitudes in the so-called multi-Regge kinematics (MRK), see e.g.~\cite{Fadin:1993wh,DelDuca:2019tur,Fadin:1975cb,Kuraev:1976ge,Kuraev:1977fs,Balitsky:1978ic}. This limit is taken such that all final-state particles are strongly ordered in rapidity, while the transverse momenta are generic. The simplest example is $2\to 3$ scattering, where the MRK limit is defined in figure \ref{fig:MRK_5points:kin}. 

One new feature compared to $2\to 2$ scattering is that signature needs to be defined separately for the target and for the projectile, hence one has four signature components in total, which we denote by ${\cal M}^{(\pm,\pm)}_{ij \to i'gj'}$.
Regge-pole factorization hold only for ${\cal M}^{(-,-)}_{ij \to i'gj'}$, and specifically, for the colour component of the amplitude where an octet representation flows across both $t$ channels (dabbed the octet-octet component). For this component the amplitude indeed factorises as 
shown in figure~\ref{fig:MRK_5points:fact} and it can be expressed as
\begin{align}
\label{pole_fact_2to3}
\left.{\cal M}^{(-,-)}_{ij \to i'gj'} \right\vert^{1\text{-Reggeon}}
&=& {\color[rgb]{0.107,0.3300,0.56250} c_i(t_1)}\,
{\olive e^{C_A \, \alpha_g(t_1)\,L_1} }\,
{\red v(t_1,t_2, {\bf p}_4^2) }\,
{\olive e^{C_A \, \alpha_g(t_2)\,L_2} }\,
{\color[rgb]{0.107,0.3300,0.56250}  c_j(t_2) }\,
{\cal M}^{\rm tree}_{ij \to i'gj'}\,.
\end{align}
Here $L_1 =\log(\frac{s_{45}}{\tau})-i\frac{\pi}{4}$
and $L_2 =\log(\frac{s_{34}}{\tau})-i\frac{\pi}{4}$
where $\tau$ is a factorization scale, which we may choose as ${\bf p}_4^2$, and ${\color[rgb]{0.107,0.3300,0.56250}  c_i(t_1) }$ and ${\color[rgb]{0.107,0.3300,0.56250}  c_j(t_2) }$ represent radiative corrections to the impact factors, $C_i=c_i C_i^{(0)}$. These impact factors and the Regge trajectory ${\olive \alpha_g(t_k)}$ are the same as in $2\to 2$ scattering, while the emission of a real gluon ($P_4$) at mid rapidity is described by the so-called \emph{Lipatov vertex}~\cite{Lipatov:1976zz,Fadin:1993wh,Fadin:1996yv,DelDuca:1998cx,DelDuca:2009ae,Fadin:2023roz}. The vertex ${\red v(t_1,t_2, {\bf p}_4^2) }=v(z,\bar{z})$ is a complex function (encoding the polarization of the emitted gluon) of a complex pair of variables, $z$ and $\bar{z}$, which are defined in terms of ratios of two-dimensional transverse momenta: $z\equiv -{\bf p}_3/{\bf p}_4$ and $1-z\equiv -{\bf p}_5/{\bf p}_4$ (defined such that momentum conservation is admitted), where
the relations with the four-dimensional momentum components are
${\bf p}_5=P_5^x+iP_5^y$ and
${\bf p}_3=P_3^x+iP_3^y$.

Currently the Lipatov vertex is known at one loop~\cite{Fadin:1993wh,Fadin:1996yv,DelDuca:1998cx,DelDuca:2009ae,Fadin:2023roz}. The availability of recent two-loop $2\to 3$ QCD amplitudes calculations~\cite{Agarwal:2023suw,DeLaurentis:2024arp,DeLaurentis:2023nss,DeLaurentis:2023izi}, opens the way to determine the Lipatov vertex at two loops. This has several independent motivations. First, it allows for a rare analytic insight into the structure of two-loop 5-point results. Second, with the Lipatov vertex at hand one can predict the MRK limit of higher-point amplitudes. Finally, the two-loop Lipatov vertex is a key ingredient in determining the BFKL kernel at the next perturbative order.
The fundamental challenge in extracting the two-loop vertex from  two-loop $2\to 3$ is the presence of multi-Reggeon exchange contributions, which much like in the case of $2\to 2$ scattering, break the simple Regge-pole factorization structure of eq.~(\ref{pole_fact_2to3}). Here we report for the first time on the computation of the multi-Reggeon exchange contributions needed to extract the Lipatov vertex. The full results for the vertex will soon be published~\cite{TBP2024}. 

\begin{figure}[htb]
\centering
\begin{subfigure}[b]{0.25\textwidth}
\centering
\includegraphics[width=0.88\textwidth]{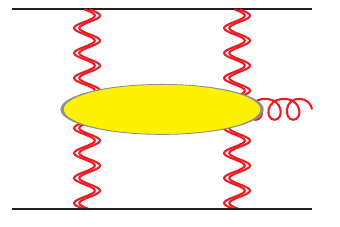}
\vspace{0em}\caption{$R^2gR^2$}
\label{R2gR2}
\end{subfigure}
\begin{subfigure}[b]{0.25\textwidth}
\centering
\includegraphics[width=0.88\textwidth]{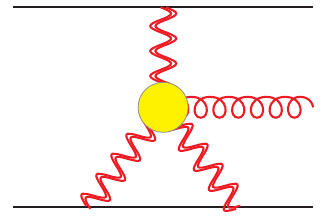}
\vspace{0em}\caption{$R^2gR$}
\label{R2gR}
\end{subfigure}
\begin{subfigure}[b]{0.25\textwidth}
\centering
\includegraphics[width=0.90\textwidth]{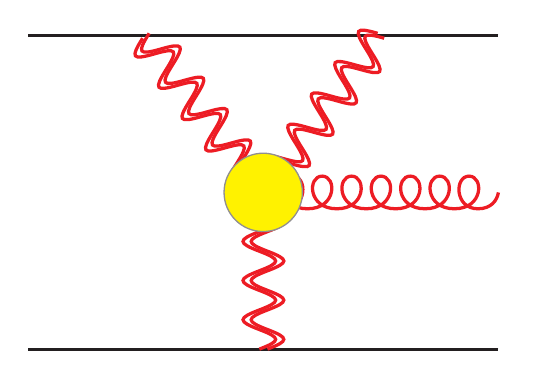}
\vspace{0em}\caption{$RgR^2$}
\label{RgR2}
\end{subfigure}
\\
\caption{Multi-Reggeon exchange at one loop in $2\to 3$ scattering}
\label{fig:MR_5points}
\end{figure}

\begin{figure}[htb]
\centering
\includegraphics[width=0.68\textwidth]{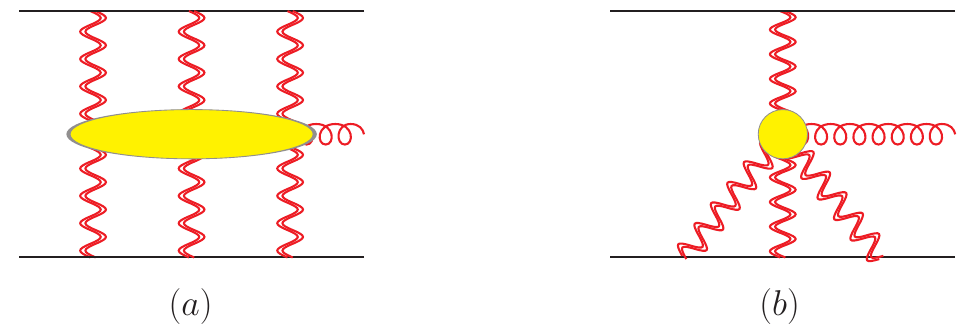}
\caption{Multi-Reggeon exchange of Odd-Odd signature at two loop in $2\to 3$ scattering}
\label{fig:MR_5points:2loops}
\end{figure}

The effective description of the high-energy limit using the shock-wave formalism discussed in section~\ref{Sec:B-JIMWLK_and_MR} can be extended to the $2\to 3$ case as follows~\cite{Caron-Huot:2013fea,Caron-Huot:2020vlo,TBP2024}.
One needs to evaluate 
\begin{align}
\label{ReggeFactBasic} 
\frac{i }{2s_{12}}
{\cal M}_{ij \to i'gj'} =
Z_i Z_j C_i^{(0)} C_j^{(0)}
\big\langle \psi_j \big| 
e^{-H L_{2}} \, a_4(p_4) \,
e^{-H L_{1}} \big| \psi_i \big\rangle, 
\end{align}
where $a_4(p_4)$ is an annihilation operator for the gluon emitted at mid rapidity with momentum $p_4$. Taking into account such an emission, requires extending the set of effective vertices as illustrated in 
figures~\ref{fig:MR_5points} and~\ref{fig:MR_5points:2loops}. At one loop one finds three possible multi-Reggeon exchange diagrams (figure~\ref{fig:MR_5points}), including a two-Reggeon exchange across the entire $t$ channel, and ones where two Reggeons are emitted from the target (projectile) side but only one Reggeon 
reaches the projectile (target). These three contribute respectively to 
${\cal M}^{(+,+)}_{ij \to i'gj'}$, 
${\cal M}^{(+,-)}_{ij \to i'gj'}$ and 
${\cal M}^{(-,+)}_{ij \to i'gj'}$,
but have no impact on the~${\cal M}^{(-,-)}_{ij \to i'gj'}$ component, and specifically, at this order the octet-octet colour flow component of the amplitude is entirely driven by a single Reggeon exchange, and admits the factorization of eq.~(\ref{pole_fact_2to3}). 

A new feature at two loops, which is illustrated in figure~\ref{fig:MR_5points:2loops}, is that the ${\cal M}^{(-,-)}_{ij \to i'gj'}$ component, and specifically, the octet-octet colour flow component of the amplitude, also receives multi-Reggeon contributions, and thus eq.~(\ref{pole_fact_2to3}) does not hold: factorization-violating terms exist. As in the case of $2\to 2$ scattering, these are expected to arise only from non-planar diagrams, so the factorization-violating terms must be subleading in the large-$N_c$ limit.

The contributions to ${\cal M}^{(-,-)\,(1)}_{ij \to i'gj'}$ at one and two loops can be summarised as follows:
\begin{eqnarray}
\label{MRcontributions_twoloops}
 {\cal M}^{(-,-)\,(1)}_{ij \to i'gj'} ={\cal M}^{(1)}_{{\cal R} g {\cal R}},\qquad\quad    {\cal M}^{(-,-)\,(2)}_{ij \to i'gj'} &=&
 {\cal M}^{(2)}_{{\cal R} g {\cal R}}
+ {\cal M}^{(2)}_{{\cal R} g {\cal R}^3} 
+ {\cal M}^{(2)}_{{\cal R}^3 g {\cal R}}
+ {\cal M}^{(2)}_{{\cal R}^3 g {\cal R}^3}\,.
\end{eqnarray}
This implies that the triple Reggeon exchange contributions in figure~\ref{fig:MR_5points:2loops} must be computed in order to extract the Lipatov vertex. Here we briefly summarise the final results of these calculations, delegating the details to~\cite{TBP2024}. 

First we note that the colour factors arising from the diagrams in figure~\ref{fig:MR_5points:2loops} (and the target-projectile symmetric counterpart) are
\begin{align}
C_{{\cal{R}}^3g{\cal{R}}^3}
    =
\begin{cases}
     \left(\frac{N_c^2}{72}+\frac12\right)\,c^{[8_a, 8_a]} -\frac{\sqrt{N_c^2-4}}{4}\,c^{[10,\overline{10}]_1} 
     & \text{for } gg
     \\
       \left(\frac{N_c^2}{72}-\frac{1}{12}+\frac{1}{4N_c^2}\right)\,c^{[8, 8]_a} 
       & \text{for } qq
        \\
       \left(\frac{N_c^2}{72}+\frac{1}{12}\right)\,c^{[8_a,8]_a} 
       & \text{for } gq
        \\
       \left(\frac{N_c^2}{72}+\frac{1}{12}\right)\,c^{[8, 8_a]_a} 
      & \text{for } qg
\end{cases}
\end{align}
\begin{align}
C_{{\cal R}g{\cal R}^3}
    =&\,
    \begin{cases}
    \left(\frac{N_c^2}{24}+\frac32\right)c^{[8_a,8_a]}
    -\frac{3\sqrt{N_c^2-4}}{4\sqrt{2}}c^{[8_a,10+\overline{10}]} & \text{for } gg
    \\
    \,
    \left(\frac{N_c^2}{24}+\frac14\right)c^{[8,8]_a} & \text{for } qq
    \\
    \,
    \left(\frac{N_c^2}{24}+\frac14\right)c^{[8_a,8]_a} & \text{for } gq
    \\
    \,
    \left(\frac{N_c^2}{24}+\frac32\right)c^{[8,8_a]_a}
    -\frac{3\sqrt{N_c^2-4}}{4\sqrt{2}}c^{[8,10+\overline{10}]} & \text{for } qg
    \end{cases}
\end{align}
The colour tensors $c^{[R_{1},R_{2}]}$ are defined in the $t$-channel colour flow basis where $R_1$ is the representation flowing into the $p_1$,$p_5$ vertex while $R_1$ into the $p_2$,$p_3$ vertex (See \cite{TBP2024} for details). As anticipated, the octet-octet component receives non-vanishing contributions for any $ij \to i'gj'$ scattering process.
Next we note that, as expected, the terms that are leading in the large-$N_c$ limit are identical for all scattering processes. This means that they can be absorbed into the Regge-pole factorized expression (\ref{pole_fact_2to3}), as expected. In contrast, the subleading terms in $N_c$ are non-universal, and hence do indeed break Regge-pole factorization.

Finally, performing the $2-2\epsilon$ dimensional loop integrals in the diagrams of figure~\ref{fig:MR_5points:2loops}, isolating the octet-octet component, and summing over the multi-Reggeon contributions in eq.~(\ref{MRcontributions_twoloops}), we find:
\begin{align}
\label{total_MR}
{\cal M}_{\text{MR}}^{(2),\,[8,8]}  =&\, \frac{(i\pi)^2}{72} 
\left(\frac{\mu^2}{|{\mathbf{p}}_4|^2}\right)^{2\eps}
{\cal M}^{(0),\,[8,8]}\,\times
    \begin{cases}
    ({\blue N_c^2}+36) {\blue F_{\text{fact}}(z,\bar{z})} &\text{for } gg
    \\
\\
    \,
     {\blue N_c^2} {\blue F_{\text{fact}}(z,\bar{z})}\,  +  \,  { F_{\text{non-fact}}^{qq}(z,\bar{z})} & \text{for } qq
    \\
\\
    \,
     {\blue N_c^2} {\blue F_{\text{fact}}(z,\bar{z})}  \, + \, { F_{\text{non-fact}}^{qg}(z,\bar{z})}   & \text{for } qg
    \end{cases}
\end{align}
where we expressed the kinematics using the complex-conjugate pair of momentum fraction variables $z$ and $\bar{z}$, defined below eq.~(\ref{pole_fact_2to3}). For the three functions defined in eq.~(\ref{total_MR}) we obtain:
\begin{align}
\label{Ffact} \nonumber
    {\blue 
F_{\text{fact}}(z,\bar{z})}&
= \frac{1}{\eps^2} - \frac1{2\eps}\log|z|^2|1-z|^2+3\,D_2(z,\bar z) - \zeta_2 + \frac54 \left(\log^2|z|^2 +\log^2|1-z|^2\right) 
\\
& - \frac12\log|z|^2\log|1-z|^2
\end{align}
while
\begin{align}
\label{Fnonfactqq}
&&F_{\text{non-fact}}^{qq} &= \,\frac{9}{\eps}\log |z|^2 |1-z|^2  +\frac{9}{2}\, \bigg(
12 D_2(z,\bar{z}) 
-\log^2 |z|^2 |1-z|^2 
\bigg)\,+
 \frac{9}{N_c^2}\bigg(\frac{1}{\eps^2} -\frac{2}{\eps} \log |z|^2 |1-z|^2  \nonumber 
 \\&&&  \qquad 
 -\,6D_2(z,\bar{z}) +2 \log^2 |z|^2 + \log |z|^2 \log|1-z|^2+2 \log^2 |1-z|^2 -\zeta_2
\bigg) 
\end{align}
\begin{align}
\label{Fnonfactqg} \nonumber 
&&\hspace*{-100pt} F_{\text{non-fact}}^{qg} &=\, \frac{27}{2\epsilon^2}-\frac{9}{\eps}\,\bigg(2 \log |z|^2 -3 \log|1-z|^2\bigg) 
+\frac{9}{2} \, \bigg(24 D_2(z,\bar{z}) 
\\
&& & \hspace*{90pt}
 +5 \log^2 |z|^2 -4 \log |z|^2 \log|1-z|^2-3\zeta_2\bigg)
\end{align}
where
\begin{equation} \label{eq:D2def}
D_2(z,\bar z) = {\rm Li}_2(z) - {\rm Li}_2(\bar z)
+ \frac{1}{2}\ln \bigg( \frac{1 - z}{1 - \bar z}\bigg) \ln (z \bar z)
\end{equation}
is the Bloch-Wigner dilogarithm. Upon subtracting the subleading terms in $N_c$ in eq.~(\ref{total_MR}), with the functions given in eqs.~(\ref{Ffact}), (\ref{Fnonfactqq}) and~(\ref {Fnonfactqg}), from the octet-octet component of the $ij\to i'gj'$  amplitude, we isolate the Regge-pole contribution. Then, using the factorization formula (\ref{pole_fact_2to3}) to divide the Regge-pole contribution by the corresponding $c_i$ and $c_j$ impact factors and Reggeized gluon factors known from $2\to 2$ scattering, we readily obtain the two-loop Lipatov vertex. Doing this for each of the three channels
$gg\to ggg$, \,$qq\to qgq$ and $qg\to qgg$ provides a robust check. 

\section{Conclusions}
\label{sec:conclusion}
We have described our research programme, aimed at understanding scattering amplitude in the (multi) Regge limit. A key element, which facilitated much of the progress reported here, is the use of iterated solutions of rapidity evolution equations to characterise and compute multi-Reggeon exchange. This framework provides an effective description of the Regge limit, entirely in terms of Reggeon degrees of freedom that live in the transverse  space.  Calculations in this framework have been done in recent years for three different applications, the signature-even NLL tower of corrections in $2\to 2$ scattering~\cite{Caron-Huot:2017zfo,Caron-Huot:2020grv,Gardi:2019pmk}, based on the BFKL equation, and the signature-odd NNLL towers in both $2\to 2$ and $2\to 3$ scattering~\cite{Caron-Huot:2017fxr,Falcioni:2020lvv,Falcioni:2021buo,Falcioni:2021dgr,TBP2024}, based on the Balitsky-JIMWLK 
equations. In this talk we focused on the signature-odd sector, where we were able to systematically separate between the factorizable Regge-pole and the non-factorizable Regge-cut contributions to partonic amplitudes. This, in combination with state-of-the-art multi-leg amplitude computations, led to major progress in determining impact factors and the gluon Regge trajectory, and soon-to-be-published, the Lipatov vertex. 

For completeness we point out that there are other approaches to describe partonic scattering in the Regge limit, which we have only mentioned here in passing. This includes a Feynman-diagram based approach to describe multi-Reggeon interactions, which has been developed and advocated in refs.~\cite{Fadin:2017nka,Fadin:2024hbe,Fadin:2023aen}, and a radically different approach based on Glauber Soft-Collinear Effective Theory~\cite{Rothstein:2016bsq,Moult:2022lfy,Gao:2024qsg}, which makes no direct use of Reggeons. There has been recent progress in these directions as well and it would surely be interesting to compare these approaches to ours.

\section*{Acknowledgments}

We would like to thank the organisers of the ``Loops and Legs in Quantum Field Theory'' conference for the opportunity to present this work.
We also wish to thank Federico Buccioni, Fabrizio Caola, Federica Devoto and Giulio Gambuti, who have been working independently on extracting the Lipatov vertex, for conducting a thorough comparison between our results.
EG is supported by the STFC Consolidated Grant ``Particle Physics at the Higgs Centre''; G.F. is supported by the EU’s Horizon Europe research and innovation programme under the Marie Sk\l{}odowska-Curie grant 101104792, \textit{QCDchallenge}.

\bibliographystyle{JHEP}
\bibliography{biblio}

\providecommand{\href}[2]{#2}\begingroup\raggedright\begin{thebibliography}{10}

\bibitem{Caola:2022dfa}
F.~Caola, A.~Chakraborty, G.~Gambuti, A.~von Manteuffel, and L.~Tancredi, {\it
  {Three-loop helicity amplitudes for quark-gluon scattering in QCD}},  {\em
  JHEP} {\bf 12} (2022) 082, [\href{http://xxx.lanl.gov/abs/2207.03503}{{\tt
  arXiv:2207.03503}}].

\bibitem{Caola:2021rqz}
F.~Caola, A.~Chakraborty, G.~Gambuti, A.~von Manteuffel, and L.~Tancredi, {\it
  {Three-loop helicity amplitudes for four-quark scattering in massless QCD}},
  {\em JHEP} {\bf 10} (2021) 206,
  [\href{http://xxx.lanl.gov/abs/2108.00055}{{\tt arXiv:2108.00055}}].

\bibitem{Caola:2021izf}
F.~Caola, A.~Chakraborty, G.~Gambuti, A.~von Manteuffel, and L.~Tancredi, {\it
  {Three-Loop Gluon Scattering in QCD and the Gluon Regge Trajectory}},  {\em
  Phys. Rev. Lett.} {\bf 128} (2022), no.~21 212001,
  [\href{http://xxx.lanl.gov/abs/2112.11097}{{\tt arXiv:2112.11097}}].

\bibitem{Agarwal:2023suw}
B.~Agarwal, F.~Buccioni, F.~Devoto, G.~Gambuti, A.~von Manteuffel, and
  L.~Tancredi, {\it {Five-parton scattering in QCD at two loops}},  {\em Phys.
  Rev. D} {\bf 109} (2024), no.~9 094025,
  [\href{http://xxx.lanl.gov/abs/2311.09870}{{\tt arXiv:2311.09870}}].

\bibitem{DeLaurentis:2024arp}
G.~De~Laurentis, {\it {Non-Planar Two-Loop Amplitudes for Five-Parton
  Scattering}},  in {\em {Loops and Legs in Quantum Field Theory}}, 6, 2024.
\newblock \href{http://xxx.lanl.gov/abs/2406.18374}{{\tt arXiv:2406.18374}}.

\bibitem{DeLaurentis:2023nss}
G.~De~Laurentis, H.~Ita, M.~Klinkert, and V.~Sotnikov, {\it {Double-virtual
  NNLO QCD corrections for five-parton scattering. I. The gluon channel}},
  {\em Phys. Rev. D} {\bf 109} (2024), no.~9 094023,
  [\href{http://xxx.lanl.gov/abs/2311.10086}{{\tt arXiv:2311.10086}}].

\bibitem{DeLaurentis:2023izi}
G.~De~Laurentis, H.~Ita, and V.~Sotnikov, {\it {Double-virtual NNLO QCD
  corrections for five-parton scattering. II. The quark channels}},  {\em Phys.
  Rev. D} {\bf 109} (2024), no.~9 094024,
  [\href{http://xxx.lanl.gov/abs/2311.18752}{{\tt arXiv:2311.18752}}].

\bibitem{Caron-Huot:2013fea}
S.~Caron-Huot, {\it {When does the gluon reggeize?}},  {\em JHEP} {\bf 05}
  (2015) 093, [\href{http://xxx.lanl.gov/abs/1309.6521}{{\tt
  arXiv:1309.6521}}].

\bibitem{Caron-Huot:2017fxr}
S.~Caron-Huot, E.~Gardi, and L.~Vernazza, {\it {Two-parton scattering in the
  high-energy limit}},  {\em JHEP} {\bf 06} (2017) 016,
  [\href{http://xxx.lanl.gov/abs/1701.05241}{{\tt arXiv:1701.05241}}].

\bibitem{Caron-Huot:2017zfo}
S.~Caron-Huot, E.~Gardi, J.~Reichel, and L.~Vernazza, {\it {Infrared
  singularities of QCD scattering amplitudes in the Regge limit to all
  orders}},  {\em JHEP} {\bf 03} (2018) 098,
  [\href{http://xxx.lanl.gov/abs/1711.04850}{{\tt arXiv:1711.04850}}].

\bibitem{Gardi:2019pmk}
E.~Gardi, S.~Caron-Huot, J.~Reichel, and L.~Vernazza, {\it {The High-Energy
  Limit of 2-to-2 Partonic Scattering Amplitudes}},  {\em PoS} {\bf RADCOR2019}
  (2019) 050, [\href{http://xxx.lanl.gov/abs/1912.10883}{{\tt
  arXiv:1912.10883}}].

\bibitem{Caron-Huot:2020grv}
S.~Caron-Huot, E.~Gardi, J.~Reichel, and L.~Vernazza, {\it {Two-parton
  scattering amplitudes in the Regge limit to high loop orders}},  {\em JHEP}
  {\bf 08} (2020) 116, [\href{http://xxx.lanl.gov/abs/2006.01267}{{\tt
  arXiv:2006.01267}}].

\bibitem{Falcioni:2020lvv}
G.~Falcioni, E.~Gardi, C.~Milloy, and L.~Vernazza, {\it {Climbing three-Reggeon
  ladders: four-loop amplitudes in the high-energy limit in full colour}},
  {\em Phys. Rev. D} {\bf 103} (2021) L111501,
  [\href{http://xxx.lanl.gov/abs/2012.00613}{{\tt arXiv:2012.00613}}].

\bibitem{Falcioni:2021buo}
G.~Falcioni, E.~Gardi, N.~Maher, C.~Milloy, and L.~Vernazza, {\it {Scattering
  amplitudes in the Regge limit and the soft anomalous dimension through four
  loops}},  {\em JHEP} {\bf 03} (2022) 053,
  [\href{http://xxx.lanl.gov/abs/2111.10664}{{\tt arXiv:2111.10664}}].

\bibitem{Falcioni:2021dgr}
G.~Falcioni, E.~Gardi, N.~Maher, C.~Milloy, and L.~Vernazza, {\it
  {Disentangling the Regge Cut and Regge Pole in Perturbative QCD}},  {\em
  Phys. Rev. Lett.} {\bf 128} (2022), no.~13 132001,
  [\href{http://xxx.lanl.gov/abs/2112.11098}{{\tt arXiv:2112.11098}}].

\bibitem{Bartels:2009vkz}
J.~Bartels, L.~N. Lipatov, and A.~Sabio~Vera, {\it {N=4 supersymmetric Yang
  Mills scattering amplitudes at high energies: The Regge cut contribution}},
  {\em Eur. Phys. J. C} {\bf 65} (2010) 587--605,
  [\href{http://xxx.lanl.gov/abs/0807.0894}{{\tt arXiv:0807.0894}}].

\bibitem{Bartels:2008ce}
J.~Bartels, L.~N. Lipatov, and A.~Sabio~Vera, {\it {BFKL Pomeron, Reggeized
  gluons and Bern-Dixon-Smirnov amplitudes}},  {\em Phys. Rev. D} {\bf 80}
  (2009) 045002, [\href{http://xxx.lanl.gov/abs/0802.2065}{{\tt
  arXiv:0802.2065}}].

\bibitem{Bern:2005iz}
Z.~Bern, L.~J. Dixon, and V.~A. Smirnov, {\it {Iteration of planar amplitudes
  in maximally supersymmetric Yang-Mills theory at three loops and beyond}},
  {\em Phys. Rev. D} {\bf 72} (2005) 085001,
  [\href{http://xxx.lanl.gov/abs/hep-th/0505205}{{\tt hep-th/0505205}}].

\bibitem{DelDuca:2019tur}
V.~Del~Duca, S.~Druc, J.~Drummond, C.~Duhr, F.~Dulat, R.~Marzucca,
  G.~Papathanasiou, and B.~Verbeek, {\it {All-order amplitudes at any
  multiplicity in the multi-Regge limit}},  {\em Phys. Rev. Lett.} {\bf 124}
  (2020), no.~16 161602, [\href{http://xxx.lanl.gov/abs/1912.00188}{{\tt
  arXiv:1912.00188}}].

\bibitem{Collins:1977jy}
P.~D.~B. Collins, {\em {An Introduction to Regge Theory and High-Energy
  Physics}}.
\newblock Cambridge Monographs on Mathematical Physics. Cambridge Univ. Press,
  Cambridge, UK, 2009.

\bibitem{White:2019ggo}
C.~D. White, {\it {Aspects of High Energy Scattering}},  {\em SciPost Phys.
  Lect. Notes} {\bf 13} (2020) 1,
  [\href{http://xxx.lanl.gov/abs/1909.05177}{{\tt arXiv:1909.05177}}].

\bibitem{Lipatov:1976zz}
L.~Lipatov, {\it {Reggeization of the Vector Meson and the Vacuum Singularity
  in Nonabelian Gauge Theories}},  {\em Sov. J. Nucl. Phys.} {\bf 23} (1976)
  338--345.

\bibitem{DelDuca:2011ae}
V.~Del~Duca, C.~Duhr, E.~Gardi, L.~Magnea, and C.~D. White, {\it {The Infrared
  structure of gauge theory amplitudes in the high-energy limit}},  {\em JHEP}
  {\bf 12} (2011) 021, [\href{http://xxx.lanl.gov/abs/1109.3581}{{\tt
  arXiv:1109.3581}}].

\bibitem{Fadin:1975cb}
V.~S. Fadin, E.~A. Kuraev, and L.~N. Lipatov, {\it {On the Pomeranchuk
  Singularity in Asymptotically Free Theories}},  {\em Phys. Lett. B} {\bf 60}
  (1975) 50--52.

\bibitem{Kuraev:1977fs}
E.~A. Kuraev, L.~N. Lipatov, and V.~S. Fadin, {\it {The Pomeranchuk Singularity
  in Nonabelian Gauge Theories}},  {\em Sov. Phys. JETP} {\bf 45} (1977)
  199--204. [Zh. Eksp. Teor. Fiz.72,377(1977)].

\bibitem{Kuraev:1976ge}
E.~A. Kuraev, L.~N. Lipatov, and V.~S. Fadin, {\it {Multi - Reggeon Processes
  in the Yang-Mills Theory}},  {\em Sov. Phys. JETP} {\bf 44} (1976) 443--450.

\bibitem{Balitsky:1978ic}
I.~I. Balitsky and L.~N. Lipatov, {\it {The Pomeranchuk Singularity in Quantum
  Chromodynamics}},  {\em Sov. J. Nucl. Phys.} {\bf 28} (1978) 822--829. [Yad.
  Fiz.28,1597(1978)].

\bibitem{Fadin:1995xg}
V.~S. Fadin, M.~Kotsky, and R.~Fiore, {\it {Gluon Reggeization in QCD in the
  next-to-leading order}},  {\em Phys. Lett. B} {\bf 359} (1995) 181--188.

\bibitem{Fadin:1995km}
V.~S. Fadin, R.~Fiore, and A.~Quartarolo, {\it {Reggeization of quark quark
  scattering amplitude in QCD}},  {\em Phys. Rev. D} {\bf 53} (1996)
  2729--2741, [\href{http://xxx.lanl.gov/abs/hep-ph/9506432}{{\tt
  hep-ph/9506432}}].

\bibitem{Fadin:2015zea}
V.~Fadin, M.~Kozlov, and A.~Reznichenko, {\it {Gluon Reggeization in Yang-Mills
  Theories}},  {\em Phys. Rev. D} {\bf 92} (2015), no.~8 085044,
  [\href{http://xxx.lanl.gov/abs/1507.00823}{{\tt arXiv:1507.00823}}].

\bibitem{DelDuca:2001gu}
V.~Del~Duca and E.~Glover, {\it {The High-energy limit of QCD at two loops}},
  {\em JHEP} {\bf 10} (2001) 035,
  [\href{http://xxx.lanl.gov/abs/hep-ph/0109028}{{\tt hep-ph/0109028}}].

\bibitem{DelDuca:2013dsa}
V.~Del~Duca, G.~Falcioni, L.~Magnea, and L.~Vernazza, {\it {Beyond Reggeization
  for two- and three-loop QCD amplitudes}},  {\em PoS} {\bf RADCOR2013} (2013)
  046, [\href{http://xxx.lanl.gov/abs/1312.5098}{{\tt arXiv:1312.5098}}].

\bibitem{DelDuca:2013ara}
V.~Del~Duca, G.~Falcioni, L.~Magnea, and L.~Vernazza, {\it {High-energy QCD
  amplitudes at two loops and beyond}},  {\em Phys. Lett. B} {\bf 732} (2014)
  233--240, [\href{http://xxx.lanl.gov/abs/1311.0304}{{\tt arXiv:1311.0304}}].

\bibitem{DelDuca:2014cya}
V.~Del~Duca, G.~Falcioni, L.~Magnea, and L.~Vernazza, {\it {Analyzing
  high-energy factorization beyond next-to-leading logarithmic accuracy}},
  {\em JHEP} {\bf 02} (2015) 029,
  [\href{http://xxx.lanl.gov/abs/1409.8330}{{\tt arXiv:1409.8330}}].

\bibitem{Fadin:2017nka}
V.~Fadin and L.~Lipatov, {\it {Reggeon cuts in QCD amplitudes with negative
  signature}},  {\em Eur. Phys. J. C} {\bf 78} (2018), no.~6 439,
  [\href{http://xxx.lanl.gov/abs/1712.09805}{{\tt arXiv:1712.09805}}].

\bibitem{Fadin:2024hbe}
V.~Fadin, {\it {Colour structure of three-reggeon cuts in QCD}},  {\em PoS}
  {\bf ICPPCRubakov2023} (2024) 037.

\bibitem{Fadin:2023aen}
V.~S. Fadin, {\it {Regge Cuts in QCD}},  {\em Phys. Part. Nucl. Lett.} {\bf 20}
  (2023), no.~3 341--346.

\bibitem{Balitsky:1995ub}
I.~Balitsky, {\it {Operator expansion for high-energy scattering}},  {\em Nucl.
  Phys.} {\bf B463} (1996) 99--160,
  [\href{http://xxx.lanl.gov/abs/hep-ph/9509348}{{\tt hep-ph/9509348}}].

\bibitem{Jalilian-Marian:1996mkd}
J.~Jalilian-Marian, A.~Kovner, L.~D. McLerran, and H.~Weigert, {\it {The
  Intrinsic glue distribution at very small x}},  {\em Phys. Rev. D} {\bf 55}
  (1997) 5414--5428, [\href{http://xxx.lanl.gov/abs/hep-ph/9606337}{{\tt
  hep-ph/9606337}}].

\bibitem{JalilianMarian:1996xn}
J.~Jalilian-Marian, A.~Kovner, L.~D. McLerran, and H.~Weigert, {\it {The
  Intrinsic glue distribution at very small x}},  {\em Phys. Rev.} {\bf D55}
  (1997) 5414--5428, [\href{http://xxx.lanl.gov/abs/hep-ph/9606337}{{\tt
  hep-ph/9606337}}].

\bibitem{JalilianMarian:1997gr}
J.~Jalilian-Marian, A.~Kovner, A.~Leonidov, and H.~Weigert, {\it {The Wilson
  renormalization group for low x physics: Towards the high density regime}},
  {\em Phys. Rev.} {\bf D59} (1998) 014014,
  [\href{http://xxx.lanl.gov/abs/hep-ph/9706377}{{\tt hep-ph/9706377}}].

\bibitem{Rothstein:2016bsq}
I.~Z. Rothstein and I.~W. Stewart, {\it {An Effective Field Theory for Forward
  Scattering and Factorization Violation}},  {\em JHEP} {\bf 08} (2016) 025,
  [\href{http://xxx.lanl.gov/abs/1601.04695}{{\tt arXiv:1601.04695}}].

\bibitem{Moult:2022lfy}
I.~Moult, S.~Raman, G.~Ridgway, and I.~W. Stewart, {\it {Anomalous dimensions
  from soft Regge constants}},  {\em JHEP} {\bf 05} (2023) 025,
  [\href{http://xxx.lanl.gov/abs/2207.02859}{{\tt arXiv:2207.02859}}].

\bibitem{Gao:2024qsg}
A.~Gao, I.~Moult, S.~Raman, G.~Ridgway, and I.~W. Stewart, {\it {A collinear
  perspective on the Regge limit}},  {\em JHEP} {\bf 05} (2024) 328,
  [\href{http://xxx.lanl.gov/abs/2401.00931}{{\tt arXiv:2401.00931}}].

\bibitem{Korchemskaya:1994qp}
I.~A. Korchemskaya and G.~P. Korchemsky, {\it {High-energy scattering in QCD
  and cross singularities of Wilson loops}},  {\em Nucl. Phys.} {\bf B437}
  (1995) 127--162, [\href{http://xxx.lanl.gov/abs/hep-ph/9409446}{{\tt
  hep-ph/9409446}}].

\bibitem{Korchemskaya:1996je}
I.~A. Korchemskaya and G.~P. Korchemsky, {\it {Evolution equation for gluon
  Regge trajectory}},  {\em Phys. Lett.} {\bf B387} (1996) 346--354,
  [\href{http://xxx.lanl.gov/abs/hep-ph/9607229}{{\tt hep-ph/9607229}}].

\bibitem{Fadin:2006bj}
V.~Fadin, R.~Fiore, M.~Kozlov, and A.~Reznichenko, {\it {Proof of the
  multi-Regge form of QCD amplitudes with gluon exchanges in the NLA}},  {\em
  Phys. Lett. B} {\bf 639} (2006) 74--81,
  [\href{http://xxx.lanl.gov/abs/hep-ph/0602006}{{\tt hep-ph/0602006}}].

\bibitem{Fadin:1993wh}
V.~S. Fadin and L.~N. Lipatov, {\it {Radiative corrections to QCD scattering
  amplitudes in a multi - Regge kinematics}},  {\em Nucl. Phys. B} {\bf 406}
  (1993) 259--292.

\bibitem{Fadin:1996yv}
V.~S. Fadin, R.~Fiore, and M.~I. Kotsky, {\it {Gribov's theorem on soft
  emission and the reggeon-reggeon - gluon vertex at small transverse
  momentum}},  {\em Phys. Lett. B} {\bf 389} (1996) 737--741,
  [\href{http://xxx.lanl.gov/abs/hep-ph/9608229}{{\tt hep-ph/9608229}}].

\bibitem{DelDuca:1998cx}
V.~Del~Duca and C.~R. Schmidt, {\it {Virtual next-to-leading corrections to the
  Lipatov vertex}},  {\em Phys. Rev. D} {\bf 59} (1999) 074004,
  [\href{http://xxx.lanl.gov/abs/hep-ph/9810215}{{\tt hep-ph/9810215}}].

\bibitem{DelDuca:2009ae}
V.~Del~Duca, C.~Duhr, and E.~W. Nigel~Glover, {\it {The Five-gluon amplitude in
  the high-energy limit}},  {\em JHEP} {\bf 12} (2009) 023,
  [\href{http://xxx.lanl.gov/abs/0905.0100}{{\tt arXiv:0905.0100}}].

\bibitem{Fadin:2023roz}
V.~S. Fadin, M.~Fucilla, and A.~Papa, {\it {One-loop Lipatov vertex in QCD with
  higher $\epsilon$-accuracy}},  \href{http://xxx.lanl.gov/abs/2302.09868}{{\tt
  arXiv:2302.09868}}.

\bibitem{TBP2024}
S.~Abreu, G.~De~Laurentis, G.~Falcioni, E.~Gardi, C.~Milloy, and L.~Vernazza,
  {\it {The Two-Loop Lipatov Vertex}},  {\em To be published} (2024).

\bibitem{Caron-Huot:2020vlo}
S.~Caron-Huot, D.~Chicherin, J.~Henn, Y.~Zhang, and S.~Zoia, {\it {Multi-Regge
  Limit of the Two-Loop Five-Point Amplitudes in $\mathcal{N} = 4$ Super
  Yang-Mills and $\mathcal{N} = 8$ Supergravity}},  {\em JHEP} {\bf 10} (2020)
  188, [\href{http://xxx.lanl.gov/abs/2003.03120}{{\tt arXiv:2003.03120}}].

\end{thebibliography}\endgroup

\end{document}